\documentclass[prl,twocolumn,aps,superscriptaddress]{revtex4}

\usepackage{graphicx}
\usepackage{amssymb,latexsym,amsthm,amsmath}
\usepackage{color}

\begin{document}

\title{Parenclitic networks: a multilayer description of heterogeneous and static data-sets}
\author{M. Zanin}
\affiliation{Faculdade de Ci\^encias e Tecnologia, Departamento de Engenharia Electrot\'ecnica,
Universidade Nova de Lisboa, Lisboa, Portugal}
\affiliation{Center for Biomedical Technology, Universidad
  Polit\'ecnica de Madrid, 28223 Pozuelo de Alarc\'on, Madrid, Spain}
\affiliation{Innaxis Foundation \& Research Institute,
Jos\'e Ortega y Gasset 20, 28006, Madrid, Spain}
\author{J. Medina Alcazar}
\affiliation{Centro de Biotecnolog\'ia y Gen\'omica de Plantas, Universidad
  Polit\'ecnica de Madrid, 28223 Pozuelo de Alarc\'on, Madrid, Spain}
\author{J. Vicente Carbajosa}
\affiliation{Centro de Biotecnolog\'ia y Gen\'omica de Plantas, Universidad
  Polit\'ecnica de Madrid, 28223 Pozuelo de Alarc\'on, Madrid, Spain}
\author{M. Gomez Paez}
\affiliation{Centro de Biotecnolog\'ia y Gen\'omica de Plantas, Universidad
  Polit\'ecnica de Madrid, 28223 Pozuelo de Alarc\'on, Madrid, Spain}
\author{D. Papo}
\affiliation{Center for Biomedical Technology, Universidad
  Polit\'ecnica de Madrid, 28223 Pozuelo de Alarc\'on, Madrid, Spain}
\author{P. Sousa}
\affiliation{Faculdade de Ci\^encias e Tecnologia, Departamento de Engenharia Electrot\'ecnica,
Universidade Nova de Lisboa, Lisboa, Portugal}
\author{E. Menasalvas}
\affiliation{Center for Biomedical Technology, Universidad
  Polit\'ecnica de Madrid, 28223 Pozuelo de Alarc\'on, Madrid, Spain}
\author{S. Boccaletti}
\affiliation{CNR- Institute of Complex Systems, Via Madonna del Piano, 10, 50019 Sesto Fiorentino, Florence, Italy}

\begin{abstract}
Describing a complex system is in many ways a problem akin to identifying an object, in that it involves defining boundaries, constituent parts and their relationships by the use of grouping laws.  Here we propose a novel method which extends the use of complex networks theory to a generalized class of "non-Gestaltic" systems, taking the form of collections of isolated, possibly heterogeneous, scalars, e.g. sets of biomedical tests. The ability of the method to unveil relevant information is illustrated for the case of gene expression in the response to osmotic stress of {\it Arabidopsis thaliana}. The most important genes turn out to be the nodes with highest centrality in appropriately reconstructed networks. The method allows predicting a set of 15 genes whose relationship with such stress was previously unknown in the literature. The validity of such predictions is
demonstrated by means of a target experiment, in which the predicted genes are one by one artificially induced, and the growth of the corresponding phenotypes turns out to feature statistically significant differences when compared to that of the wild-type.

\end{abstract}

\maketitle

Of the different ways of representing a multi-unit system, the one afforded
by complex networks is among the most elegant and general. In the last
years, complex networks \cite{Albert2002,Barabasi2004} have provided a
valuable framework for the analysis of a wealth of natural and man-made
systems, in fields as diverse as, amongst others, genetics, proteomics and
metabolomics \cite{Barabasi2004}, the study of neurological diseases \cite%
{Bullmore2009}, transportation networks \cite{Zanin2013} and the World Wide
Web \cite{Albert1999}.
While graph theory allows characterizing systems as soon as they are
identified as an object, it says nothing as to what should be treated as
such. Defining boundaries and identifying components of a complex system can
be a natural, if not trivial, task as in the case of an ensemble of power
grids, for which it is immediately evident what nodes, links and system
boundaries are. In other cases, where individual components are
well-defined, some relationship, either conceptual, semantic, or functional,
e.g. friendship in a social network \cite{Wassermann1994} or correlated
activity at different brain regions \cite{Bullmore2009,Bassett2006}, helps
segmenting the system from its surroundings.

Often, however, defining what can indeed be treated as a system in the first
place, may be highly non trivial. Suppose, for instance, that what one wants
to study is a set of biomedical data from different individuals, e.g.
various blood tests, which are in essence but a collection of scalar values
without any history. \textit{Prima facie}, such an object study would seem
to lack the physical or virtual relationships between elements of the
system, which anatomic brain fibres or hyper-links respectively provide for
brain tissue and pages of a web site. Nor does it appear to be possible to
construct the sort of functional links that one can define when time
evolving variables are associated to each node, as e.g. the time evolution
of a stock price, or of brain activity in a given region. Therefore, whether
and how such a matter should be treated as a unitary system is not obvious.
In particular, what would the elements be of such a system and how would
internal relationships among them be defined?

\begin{figure}[h!]
\includegraphics[width=0.45\textwidth]{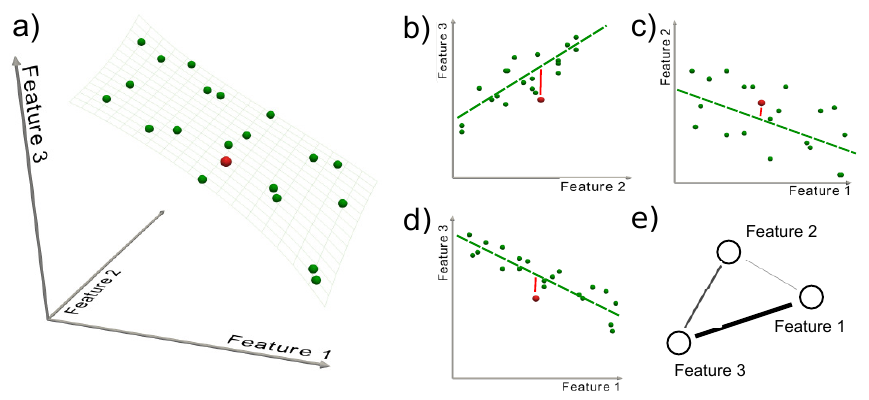}
\caption{(Color online) {\it Schematic illustration of the parenclitic network reconstruction method}.
(a) The initial data set, for three features, corresponds to a set of points (green spheres) in a 3-dimensional space. The constraint surface (gray wired surface) represents the overall standard relationship of the class. A generic unlabeled subject is represented by a red sphere. (b,c,d) Data are then projected on each of the three possible planes. The green dashed lines represent the models extracted in each plane. The red points are the positions of the unlabeled subject, and the red lines indicate the distance of the subject from the models. (e) The resulting parenclitic representation is a network where nodes are associated to features, and links are weighted according to the calculated distances (coded, in this Figure, into different line widths).
\label{fig:1}}
\end{figure}

In this Letter, we introduce a novel way of representing collections of
isolated, possibly heterogeneous, scalars as complex networks, wherein the
constitutive elements (the nodes) are a group of features characterizing a
given subject, and links are weighted according to the deviation between the
values of two features and their corresponding typical relationship within a
studied population. The result is what we term here a \textit{parenclitic}
network representation, from $\pi \alpha \rho \acute{\varepsilon}\gamma
\kappa \lambda \iota \sigma \iota \varsigma $, the Greek term for
"deviation", originally used by the Greek philosopher Epicurus to designate
the spontaneous and unpredictable swerving of free-falling atoms, allowing
them to collide \cite{Lucretius}.

The starting point is a multi-feature description of subjects, e.g. a
collection of medical measurements or of genetic expression levels, and a
subjects' affiliation to one or multiple predefined groups. While working
with the complete data set may result unfeasible, we consider the projection
of the data into all possible plains created by pairs of features. In these
plains, different methods (from simple linear correlations, up to more
sophisticated data mining techniques) are used to extract reference models
for each group. When a new, unlabeled, subject is considered, the deviation
between the associated data and such reference models is used to weight the
link between the corresponding nodes.

In general terms, consider a set of $n$ systems, or subjects, $\{s_1, s_2,
\ldots, s_n \}$, each one associated to one of $n_c$ pre-defined classes -
the class of each system will be denoted by $\{ c_1, c_2, \ldots, c_{n_c} \}$%
. For instance, each subject may represent a person, classified as \textit{%
healthy} (or \textit{control}) or suffering from some disease. Each subject $%
i$ is, in turn, identified by a vector of $n_f$ features $f_i = (f ^i _1, f
^i _2, \ldots, f ^i _{n_f})$, so that each system is represented by a point
in a $n_f$-dimensional space.

\begin{figure}[h!]
\includegraphics[width=0.45\textwidth]{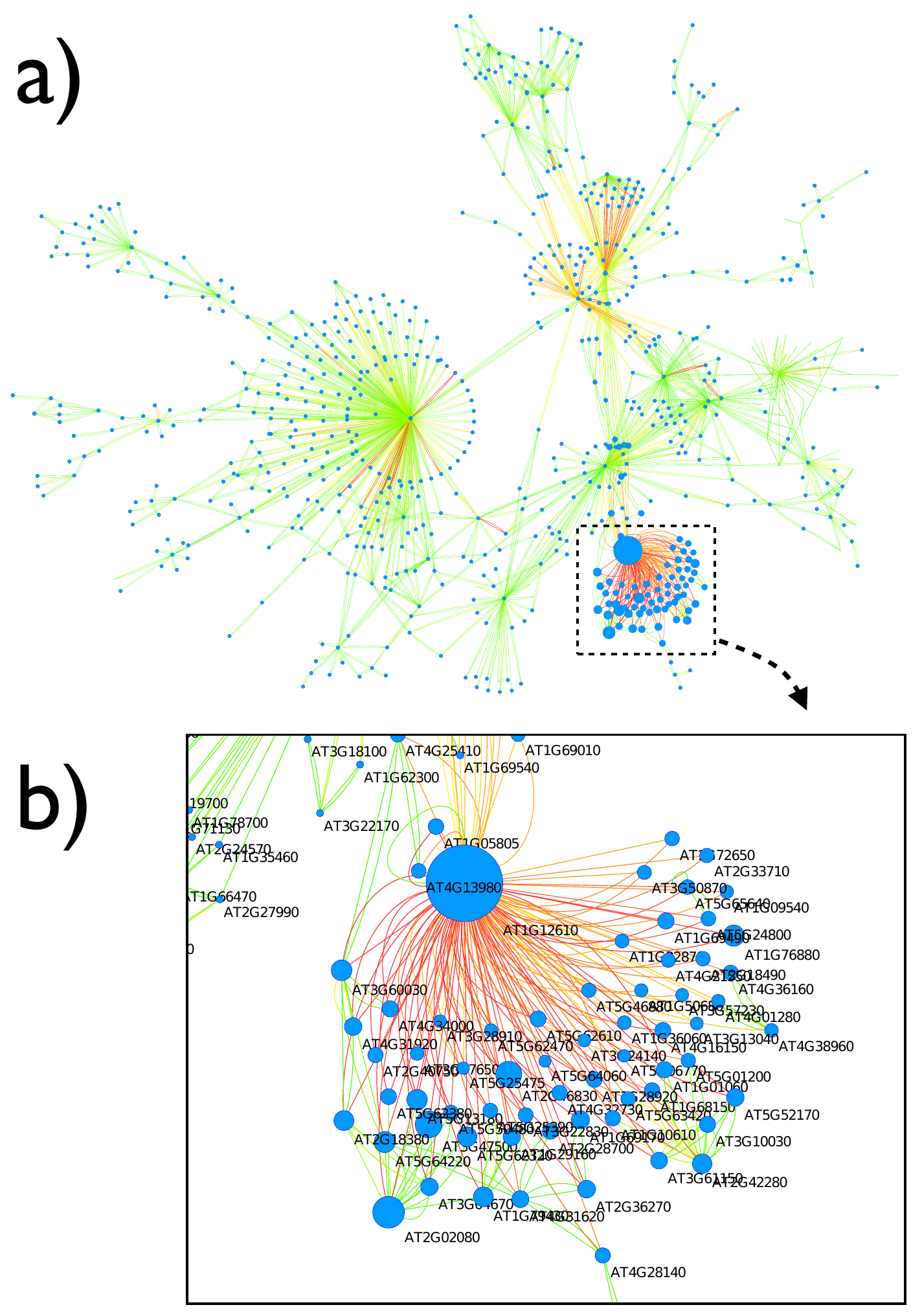}
\caption{(Color online) {\it Parenclitic network for the response of Arabidopsis thaliana to osmotic stress after 3 h.}
(a) Pictorial representation of the resulting parenclitic network; for the sake of clarity, links with weight lower than $7$ are not depicted. (b) Magnification of the neighborhood of the most central node, {\it AT1G12610}. In both cases, color represents the link weight (from green to red), and node size is associated with the corresponding value of $\alpha$-centrality.
\label{fig:2}}
\end{figure}

The fundamental ansatz is that each class can be associated to a set of
constraints in the features' space. In other words, for each pairs of
features $i$ and $j$, the values corresponding to subjects of a given class $%
c$ are supposed to lie on a constraint $\tilde {\mathcal{F}} _{i,j} ^c (f_i,
f_j)=0$, modeling the relationship expected in that plane for subjects
belonging to that class (see Fig. \ref{fig:1} for a schematic illustration).
Such reference models can be obtained by several methods, like for instance
a polynomial fit, or more generally by a data mining method like Support
Vector Machine or Artificial Neural Networks. For each unlabeled subject, a
parenclitic network of $n_f$ nodes is then constructed where vertices
represent the features, and the distance between the subject's position in
the plane of features $i$ and $j$ and the derived model $\tilde {\mathcal{F}}
_{i,j} ^c (f_i, f_j)=0$ is used to weight the link between nodes $i$ and $j$
- see the red dot and line in Fig. \ref{fig:1} (b) and the resulting
topology illustrated in Fig. \ref{fig:1} (c). Notice that each of the
ensuing $\frac{n_f (n_f-1)}{2}$ links is here a vector of $c$ scalar
components, and therefore the arising network is intrinsically a \textit{%
multilayer} network, where each layer quantifies the subject deviation from
one of the pre-defined classes. Suitable, optimized, thresholding techniques
can be used to later transform such a weighted clique into a structured,
sparse, network \cite{zaninzanin}.

The topological characteristics of the parenclitic network can then be used
to extract relevant, otherwise inaccessible, information about the system.
In particular, atypical or pathological conditions correspond to strongly
heterogeneous networks, whereas typical or normative conditions are
characterized by sparsely connected networks with homogeneous nodes \cite%
{Zanin2011}. Insofar as a network representation of each instance is
constructed with reference to the population to which it is compared, this
technique is by its very nature a difference seeker.

While such a graph representation is of general applicability to all systems
whose available information is limited to collections of static expressions
of features (and it also allows merging different data sources into a single
network), in the following we will illustrate its details and prediction
power in a specific, relevant, context: the genetic expression of the plant
\textit{Arabidopsis thaliana} under osmotic stress, with the objective of
identifying those genes orchestrating the plant's response under such a
stress condition.

\begin{figure}[h!]
\includegraphics[width=0.45\textwidth]{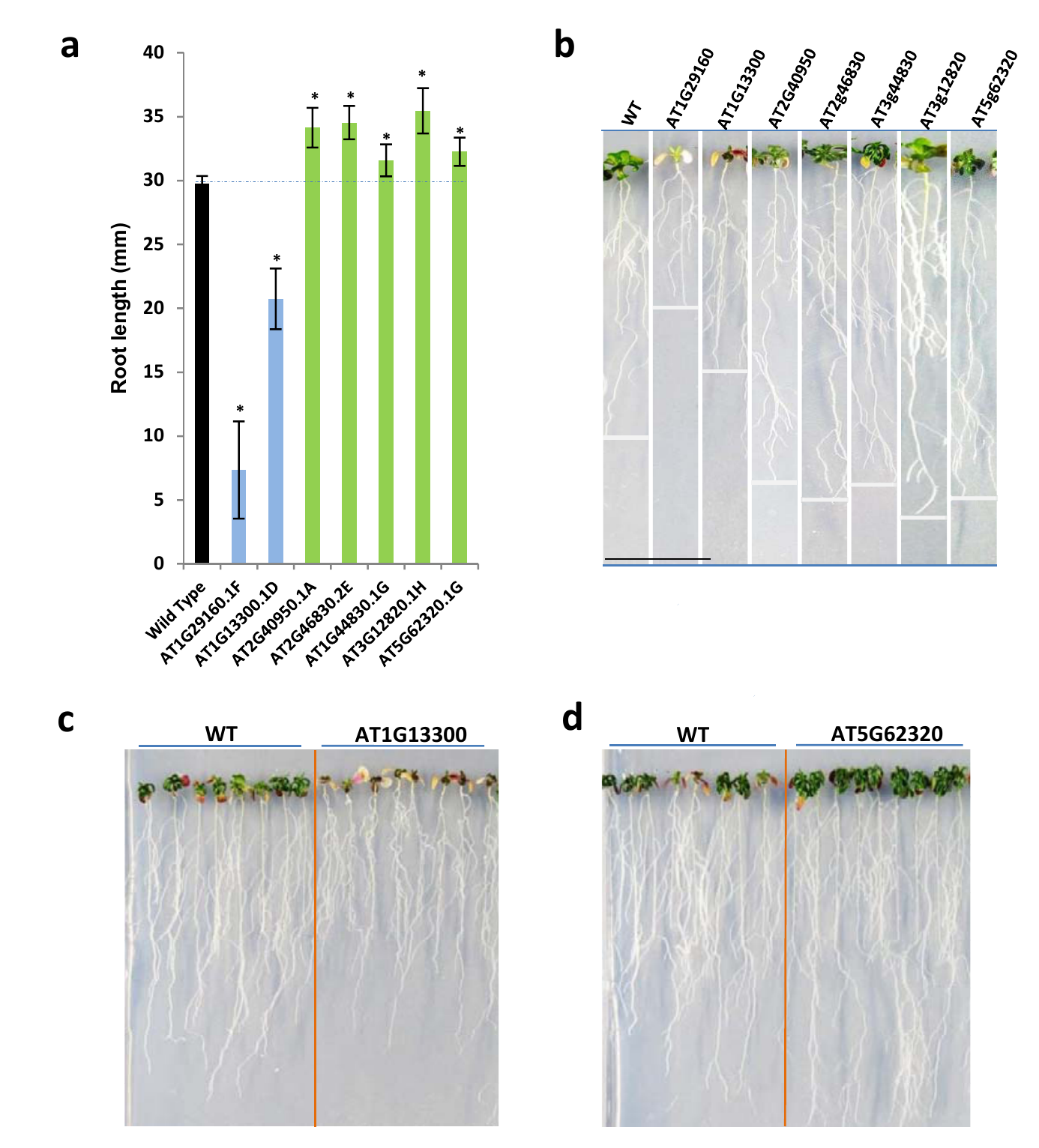}
\caption{(Color online) {\it In vivo experimental verification of the predictions.}
(a) Mean root length corresponding to the wild type (WT, black column) and to 7 other transgenic lines in which a specific gene has been knocked out. Whiskers represent the standard deviation corresponding to each group. Asterisks denote groups for which the distribution of root lengths is different with respect to the wild type with a $0.01$ significance level. (b) Photos of one plant of each of the 8 lines, at the end of the full development process. (c) and (d) Photos of two vertical plates where plants are grown. In both cases, the left (right) photos refer to wild phenotypes (to phenotypes developed by the transgenic line).
\label{fig:3}}
\end{figure}

Data are obtained from the \textit{AtGenExpress project} \cite{Kilian2007},
including expression levels of $22,620$ genes under 8 different abiotic
stresses (i.e., cold, heat, drought, osmotic, salt, wounding and UV-B light)
and at six different moments of time (30 min, 1 h, 3 h, 6 h, 12 h and 24 h
after the onset of stress treatment). Of these, we focus in the following
only on the osmotic stress, and the analysis is then performed onto the $%
n_f=1,922$ genes composing the transcription factors of Arabidopsis \cite%
{Guo2005}. While the classical approach considers co-expression networks
\cite{Clifton2005}, the parenclitic network representation focuses on those
pairs of genes whose expressions depart from a reference model. The two
methods are therefore strongly complementary: the former focusing on
similarities between the evolutions of expression levels through time, the
latter concentrating on differences.

In our approach, we create a network for each time step by considering as
"subjects" the statuses of the plant at the other time steps, this way
concentrating on those pairs of features whose current relationship deviates
from that of all other times. In other words, when analyzing data at time $%
\tau$, we create the $n_f(n_f-1)$ reference models $\{ \tilde{\mathcal{F}}=0
\}$ for the unique class ($c=1$) of those data corresponding to all other
time steps, and we generate links according to the distance from that
reference.

\begin{figure}[h!]
\includegraphics[width=0.4\textwidth]{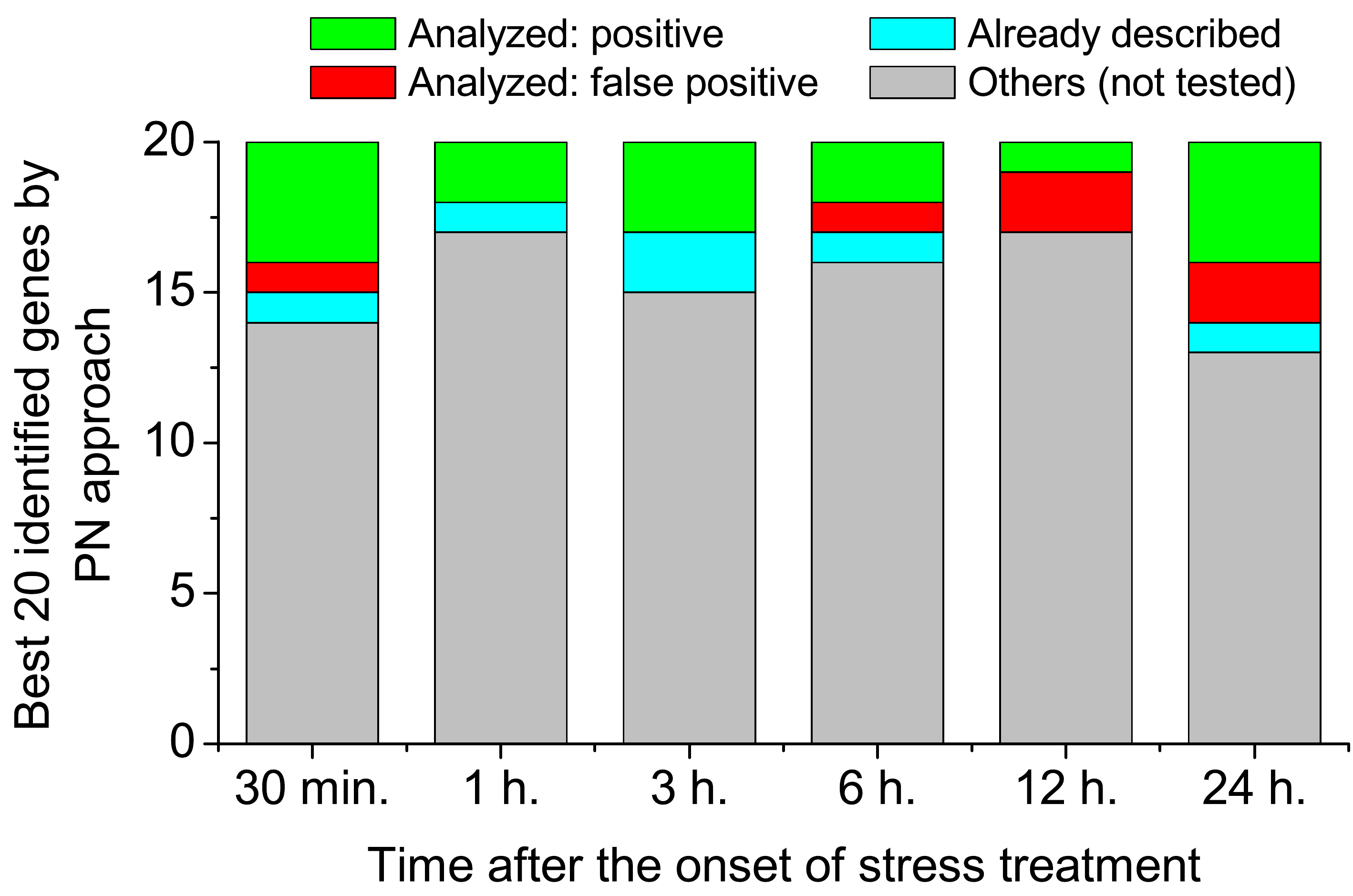}
\caption{(Color online) {\it Screening of the experimental results.}
Bars account for the 20 most central genes at each time step. For the six time steps considered, bar colors are coded according to the following stipulations: genes previously considered not to be involved in the plant's response to osmotic stress, that were respectively experimentally proven to develop (green) or to fail to develop (red) a statistically significant difference in the phenotype with respect to the wild phenotype; (cyan) genes predicted by the parenclitic analysis that were previously associated with the stress response in the Literature; and (gray) previously unknown genes, which could not be tested experimentally.
\label{fig:4}}
\end{figure}

Precisely, given two gene expression levels $i$ and $j$, we define our
reference models by linear regression as $\tilde{ f _j ^\tau } = \alpha
_{ij} + \beta _{ij} f _i ^{\tau}$, where $\tilde{ f _j ^\tau }$ is the
expected value of gene $j$ at time $\tau$, $f _i ^{\tau}$ the known
expression levels of gene $i$, and $\alpha _{ij}$ and $\beta _{ij}$ two free
model parameters. These two coefficients are calculated by means of a linear
fit of all values corresponding to other time steps, i.e., minimizing the
error of the relation $f _j ^{t \ne \tau} = \tilde{\mathcal{F}}(f _i ^{t \ne
\tau}) = \alpha _{ij} + \beta _{ij} f _i ^{t \ne \tau}$. The distance
between the expected (corresponding to the model $\tilde{\mathcal{F}}(f _i
^{t \ne \tau})$) and the real value of gene $j$ is then used to weight the
link connecting nodes $i$ and $j$ in the network. More specifically, the
weight of the link is the absolute value of the Z-Score of the distance $%
\left| {\ \tilde{ f _j ^\tau } - f _j ^\tau } \right|$.

As for the identification of the more central nodes (i.e., genes) within
each of the six parenclitic networks, we opted for the $\alpha- centrality$
measure, according to which the centrality of a node is a linear combination
of the centralities of those to whom it is connected \cite{Bonacich2001}. If
we define a vector $\mathcal{X}$ of centralities such that its $i^{th}$
component $x_i$ is the centrality of the $i$-th node, we have $\lambda x_i =
\sum\limits_j {x_j (W_{ij} + \alpha) }, \ \ (W + \alpha) \mathcal{X} =
\lambda \mathcal{X}$. Here, $W$ is the weight matrix of the network, and $%
W_{i,j}$ codifies the weight of the link connecting nodes $i$ and $j$.
Notice that this is equivalent to an eigenvalue problem, with constant $%
\alpha$ defining weak connections between all the nodes of the network. In
order to have meaningful results, $\alpha$ should be smaller than the
spectral radius of $W$.

An example of the obtained networks is shown in Fig. \ref{fig:2}. Namely,
Fig. \ref{fig:2} (a) depicts the giant component of the network \cite{footnote}
corresponding to 3 h. The color of links accounts for their weights, with
green (red) shades indicating low (high) Z-Scores, and the size of nodes is
proportional to their $\alpha-centrality$. Remarkably,
the resulting network topologies are characterized by a high heterogeneous
structure, dominated by a small number of \textit{hubs}  \cite{footnote} - as can be
appreciated from the zoom reported in Fig. \ref{fig:2} (b). Such highly
central nodes indicate that, at 3 h., the expression levels of the
corresponding genes strongly deviate from the relationships generally
established at other times. This suggests that such genes are performing some
specific task at this time point, and therefore that they are the main
actors in regulating the overall plant response. Thanks to this parenclitic
network representation, $15$ new genes, either previously unknown or
considered unrelated to the response to osmotic stress, were identified \cite%
{foot}, the full list of which is reported in Table \ref{tab:NewGenes}.

\begin{table}
  \begin{tabular}{| l | l | l | }
    \hline
	Time step & Gene &  Centrality \\ \hline
	
	30 m. & AT1G13300 & 0.88111 \\
	30 m. & AT5G51910 & 0.729679 \\
	30 m. & AT4G23750 & 0.507826 \\ \hline
	1 h. & AT1G44830 & 1.0 \\
	1 h. & AT3G12820 & 0.236686 \\ \hline
	3 h. & AT2G46830 & 0.271497 \\
	3 h. & AT5G62320 & 0.177404 \\
	3 h. & AT1G29160 & 0.148112 \\ \hline
	6 h. & AT4G16610 & 0.767785 \\
	6 h. & AT2G44910 & 0.689358 \\ \hline
	12 h. & AT3G61910 & 0.264721 \\ \hline
	24 h. & AT1G09540 & 0.709785 \\
	24 h. & AT2G40950 & 0.551008 \\
	24 h. & AT5G62320 & 0.482752 \\
	24 h. & AT5G04410 & 0.438538 \\
    \hline
  \end{tabular}
	\caption{\label{tab:NewGenes} {\it Genes previously unknown in the Literature, discovered by the parenclitic network representation, and experimentally proven to develop a statistically significant phenotype.} The right most column reports the corresponding centrality values, as normalized to that of the most central node of the 6 networks (the
gene AT1G44830).}
\end{table}

To confirm these predictions, we further performed an \textit{in vivo}
experiment, in which genes corresponding to the most central nodes of each
graph were knocked out, and the appearance of some phenotype was monitored
by measuring the length of the root of each plant. Precisely, for the
screening of the transcription factors identified by the parenclitic model,
the \textit{Arabidopsis thaliana} inducible lines from Transplanta
collection \cite{transplanta} were used, with the ecotype Columbia (Col-0)
as the Wild Type. Each one of the transgenic \textit{Arabidopsis} lines of
the collection expresses a single \textit{Arabidopsis} transcription factor
under the control of the $\beta$-stradiol inducible promoter. In the
experiment, seeds from control plants (Col-0) and at least two independent
T3 homozygous transgenic lines (Transplanta collection \cite{transplanta})
of each transcription factor were sterilized, vernalized for 2 days at $%
4^\circ$C and plated onto Petri dishes containing $\frac{1}{2}$ MS medium
\cite{Murashige1962} supplemented with $10 \mu$M $\beta$-Stradiol. After 5
days, seedlings were transferred to vertical plates containing $\frac{1}{2}$
MS medium supplemented with 300 mM Mannitol, $10 \mu$M $\beta$-stradiol and
transferred to a growth chamber at $21^\circ$C under long-day growth
conditions (16/8h light/darkness). After 12 days pictures were taken to
record the phenotypes, and root elongation measurements were performed with
ImageJ software \cite{Abramoff2004}.

As an example, Fig. \ref{fig:3} reports the results obtained with seven
transgenic lines, i.e. seven groups of plants in which the expression of one
gene was artificially suppressed. Specifically, Fig. \ref{fig:3} (a) reports
the mean length of roots for the seven lines, as compared to the expected
root length in the wild type (i.e., the plant without genetic modifications,
black column). The Figure clearly visualizes the fact that, in all the seven
examples, knocking down the corresponding gene leads to a strongly abnormal
development of the plant. The complete results of the \textit{in vivo}
screening are summarized in Fig. \ref{fig:4}. For each of the six networks
analyzed, Fig. \ref{fig:4} reports the number of genes already known to be
relevant for the osmotic response of the plant, and the number of previously
unknown genes that have been successfully tested.

In conclusion, the parenclitic approach allows a network representation of
those data sets lacking both a physical background of connections, and a
time-varying nature. Yet, by exploiting the data associated to a set of
pre-labeled subjects, and by extracting a set of reference models, it is
possible to construct networks whose links represent the presence of
deviations from expected relationships. This representation unveils
important information on the system, as the identification of key genes
regulating the response of the plant \textit{Arabidopsis thaliana} to
osmotic stress, whose role was previously unknown in the literature. Our
method generalizes network representation to a very vast number of contexts
and data sets previously thought to be outside graph theory's domain of
application.

Authors acknowledge Shlomo Havlin for many fruitful discussion on the
subject, as well as the computational resources and assistance provided by
CRESCO, the center of ENEA in Portici, Italy.


\end{document}